\documentclass[aps,prb,reprint,showpacs]{revtex4-1}

\usepackage{amsmath}
\usepackage{graphicx}
\usepackage{hyperref}

\usepackage{xcolor}

\usepackage{soul}

\begin{document}

\title{Magnetic polaron and antiferro-ferromagnetic transition in doped bilayer CrI$_3$}

\author{D. Soriano$^{\rm{1}}$, M. I. Katsnelson$^{\rm{1}}$}

\affiliation{$^{\rm{1}}$Radboud University, Institute for Molecules and Materials, NL-6525 AJ Nijmegen, the Netherlands}

\begin{abstract}
Gate-induced magnetic switching in bilayer CrI$_3$ has opened new ways for the design of novel low-power magnetic memories based on van der Waals heterostructures. The proposed switching mechanism seems to be fully dominated by electrostatic doping. Here we explain, by first-principle calculations, the ferromagnetic transition in doped bilayer CrI$_3$. For the case of a very small electron doping, our calculations predict the formation of magnetic polarons (``ferrons'', ``fluctuons'') where the electron is self-locked in a ferromagnetic droplet in an antiferromagnetic insulating matrix. The self-trapping of holes is impossible, at least, within our approximation.
\end{abstract}

\maketitle


{\textbf{\textit{Introduction.}}} The discovery of long-range ferromagnetic order in two-dimensional semiconductors\cite{HuangXu2017,GongZhang2017,DengZhang2018,GibertiniNovoselov2019} has opened new venues for the design and engineer of novel magneto-optic\cite{ZhongXu2017,SeylerXu2018}, magneto-electronic\cite{ZollnerFabian2017,SongXu2018,KleinJarillo2018,WangMorpurgo2018,CardosoRossier2018,GhazaryanMisra2018,WangMorpurgo2019,RustagiUpadhyaya2019} and spintronic\cite{KarpiakDash2019,Cummings2019,KimGeim2019} devices based on van der Waals heterostructures. In chromium trihalides (CrX$_3$), most of these applications rely on their layered-antiferromagnetic ground state, and on the low critical fields needed for a ferromagnetic phase transition (0.6-0.7 T and 1.1 T for bilayer CrI$_3$ and CrCl$_3$ respectively). Recent experiments by Thiel {\it et al.}\cite{ThielMaletinsky2019} and Ubrig {et al.}\cite{UbrigGibertini2019} have confirmed that the observed antiferromagnetic ground-state in CrI$_3$ is related to the different layer stacking in bulk and few-layer samples. These experimental observations are in agreement with previous first-principles calculations predicting a strong reduction of interlayer exchange in bilayer CrI$_3$ when going from rhombohedral to monoclinic stacking.\cite{SivadasXiao2018,JiangJi2018,JangHan2018,SorianoRossier2019}

\begin{figure}[t]
	\centering
	\includegraphics[clip=true, width=0.45\textwidth] {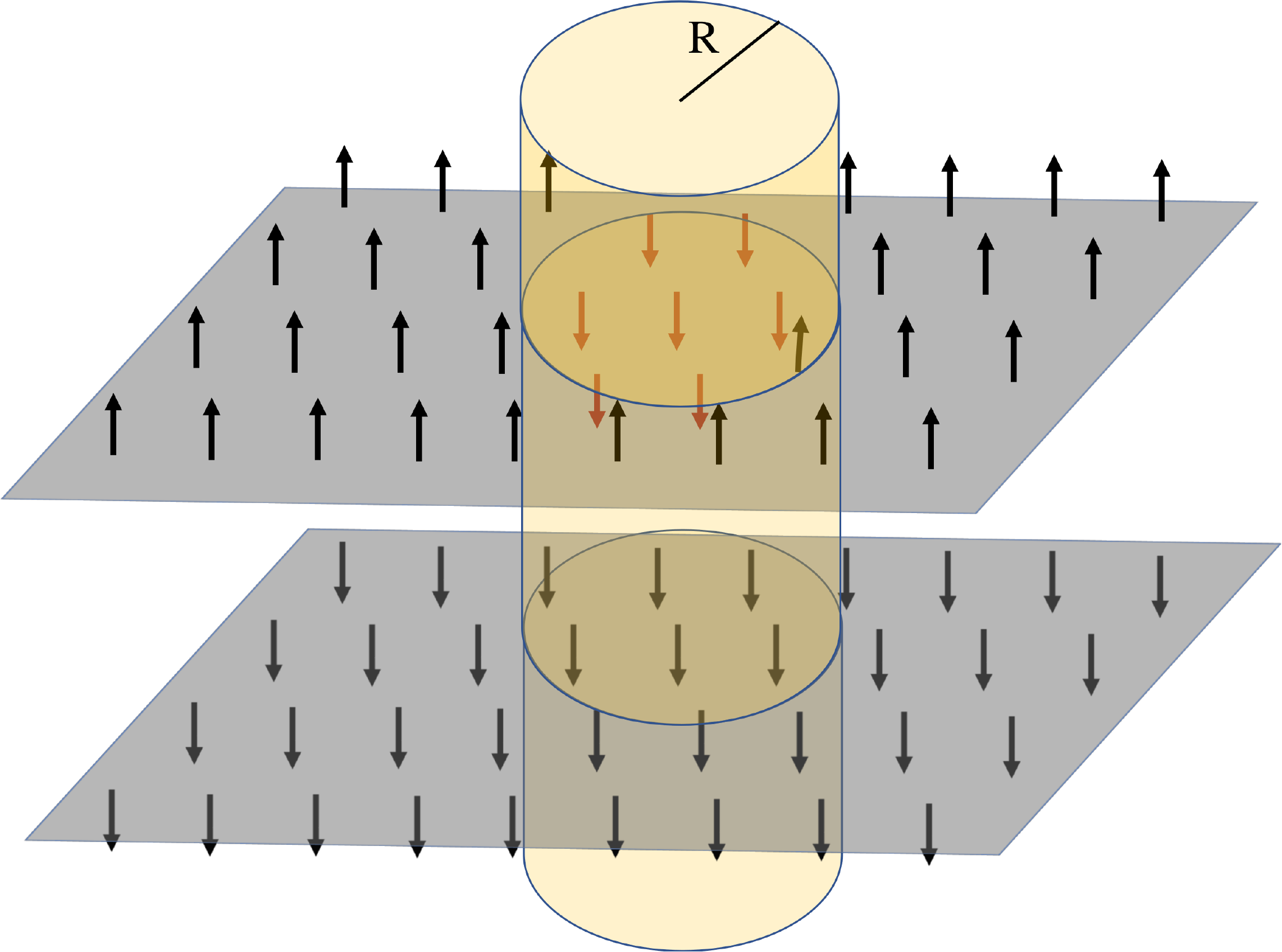}
	\caption{\label{FIG1} Schematic view of a magnetic polaron (yellow area) in a bilayer antiferromagnetic system.}
\end{figure}

The low critical fields reported for bilayer CrI$_3$ denote a weak interlayer exchange coupling able to be affected by other external perturbations. Recent experiments on dual-gated bilayer CrI$_3$ have demonstrated that an electron doping of $n \approx 2\times10^{13}$ is able to switch the interlayer exchange coupling from antiferromagnetic(AFM) to ferromagnetic(FM)\cite{JiangShan2018}. A similar behaviour has been reported for bilayer and few-layer CrI$_3$ close to the interlayer spin-flip transition\cite{JiangMak2018,HuangXu2018}, and also for Cr$_2$Ge$_2$Te$_3$ samples,\cite{WangZhang2018} thus allowing for a fully electrical control of interlayer magnetism in few-layer CrI$_3$. A plausible mechanism describing magnetic transitions through electrostatic doping is based on the formation of magnetic polarons, as shown schematically in Fig.\ref{FIG1}, where a self-trapped electron is forming a local ferromagnetic environment of radius ($R$) in a bilayer antiferromagnetic system.

In this work, we shed some light on the mechanism underlying interlayer magnetic transition in electron/hole doped bilayer CrI$_3$ by combining first-principles calculations and an effective low-energy model that allows us to describe the energy of the magnetic polaron in terms of the interlayer exchange coupling and the carrier bandwidth. In the first part, we perform {\it ab-initio} calculations in doped bilayer CrI$_3$ and show the effect of electrostatic doping on the interlayer exchange coupling. In the second part, we introduce the effective model explaining qualitatively our numerical results via the formation of magnetic polarons in layered semiconducting magnets. In the last part, we discuss and summarize the results.

\smallskip

\begin{figure}[t]
\centering
\includegraphics[clip=true, width=0.45\textwidth] {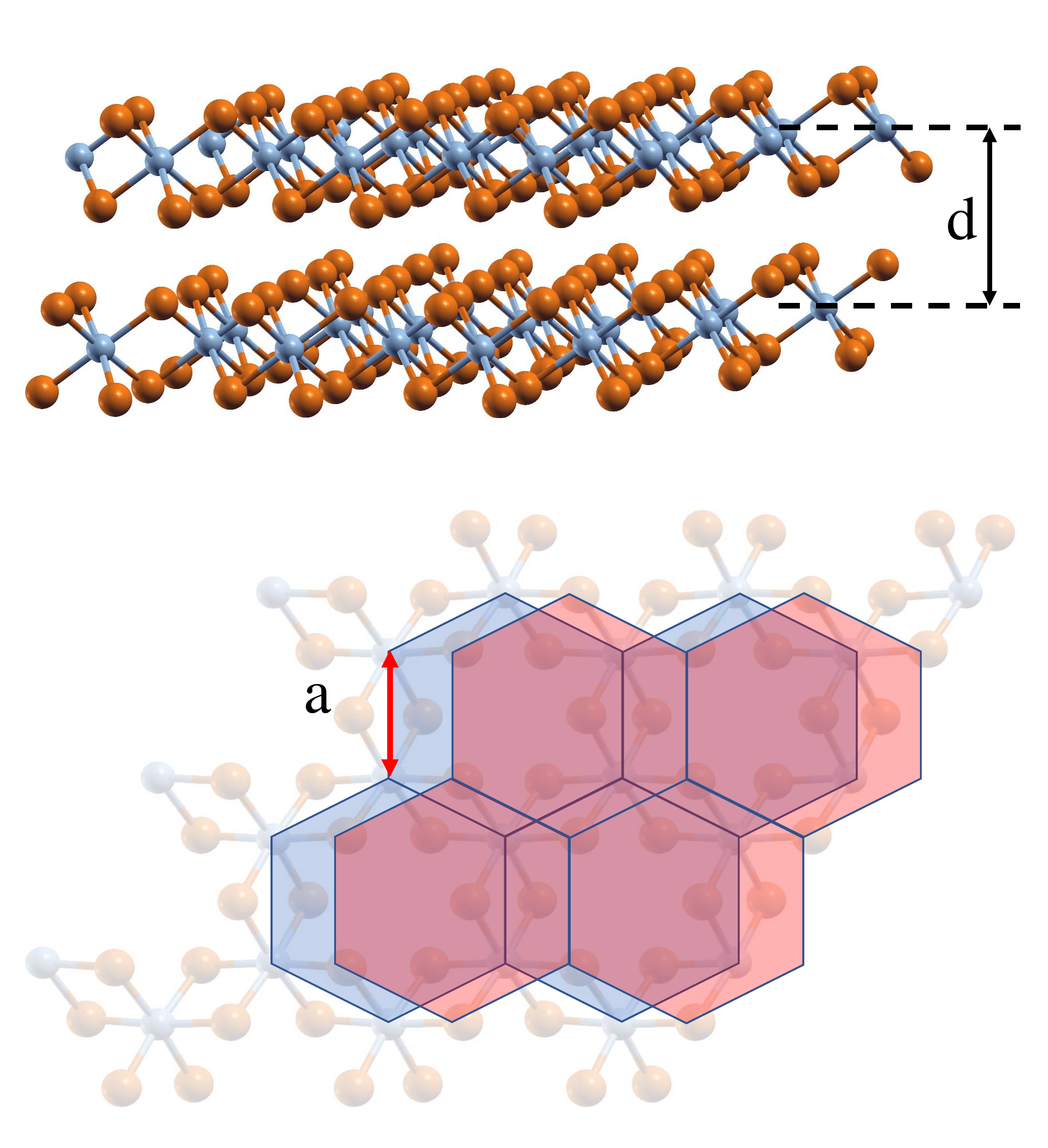}
\caption{\label{FIG2} Detail of the crystal structure of bilayer CrI$_3$. The intralayer ($a$) and interlayer ($d$) Cr-Cr distances are highlighted in the bottom and top panels respectively. In each layer, Cr atoms (in blue) form an hexagonal lattice sandwiched by two layers of I atoms (orange). In the bottom pannel, we show a scheme of the monoclinic stacking. Blue and red hexagons represent different layers. }
\end{figure}

{\textbf{\textit{First-principles calculations.}}}  We have performed density functional theory (DFT+U) calculations  on monoclinic bilayer CrI$_3$ (see Fig.\ref{FIG2}) using the plane-wave based code PWscf as implemented in the Quantum-Espresso \textit{ab-initio} package\cite{QE}. The Cr-Cr intralayer and interlayer distances obtained after relaxation are $a = 3.9647$ \AA$~$ and $d = 6.6213$ \AA. The quasi-Newton algorithm for ion relaxation is applied until the components of all forces are smaller than $10^{-3}$ Ry. For the self consistent calculations, we use $U = 3$ eV and a $8 \times 8 \times 1$ $k$-point grid. Projector augmented wave (PAW) pseudopotentials within the Perdew-Burke-Ernzerhoff (PBE) approximation\cite{perdew96} for the exchange-correlation functional are used for Cr and I atoms. Van der Waals dipolar corrections are introduced through the semiempirical Grimme-D2 potential.\cite{grimme06} Spin-orbit interactions are not included in our calculations. 

\begin{figure}[t!]
	\centering
	\includegraphics[clip=true, width=0.48\textwidth] {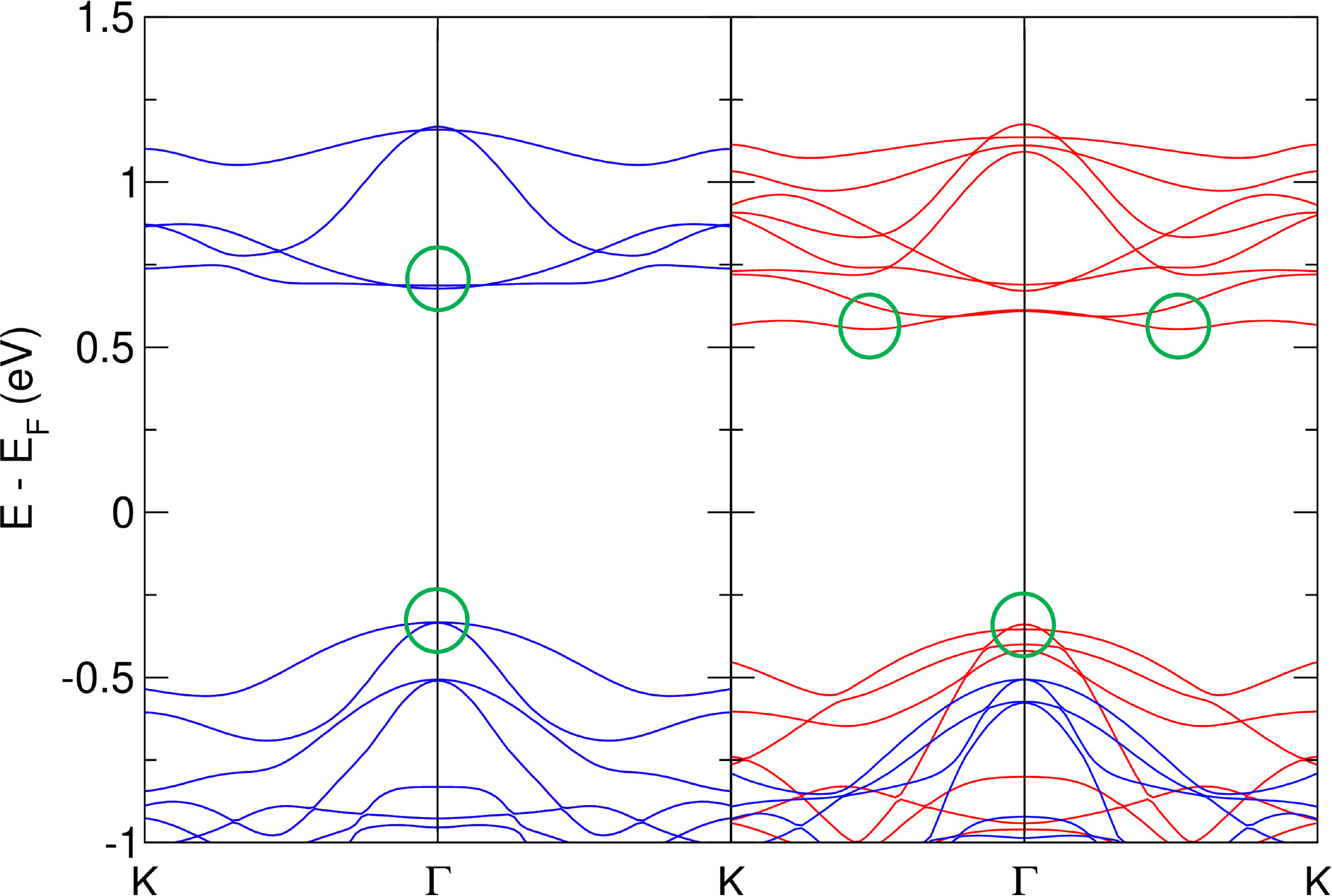}
	\caption{\label{FIG3} Band structure of monoclinic bilayer CrI$_3$ for AFM (left) and FM (right) interlayer exchange. The green circles denote the maximum and minimum of the valence and conduction bands respectively.}
\end{figure}

In Fig.\ref{FIG3}, we show the $\Gamma$-centered band structure of  monoclinic bilayer CrI$_3$ for AFM and FM interlayer coupling. In the FM configuration, the strong hybridization of the empty Cr $e_g$-orbitals localized on different layers modifies the dispersion of the conduction bands. This results in a lower and indirect bandgap  in the FM case compared to the AFM one (green circles in Fig.\ref{FIG3} indicate the top and bottom of conduction and valence bands). In the first two columns of Table \ref{TAB1}, we show the energy values of the bottom(top) of conduction(valence) bands for AFM and FM interlayer coupling, as well as the calculated effective masses ($m^*$) for electron and hole carriers.

\begin{table}[t]
	\caption{Summary of the parameters obtained from first-principles ($m_e$ is the free electron mass)}	
	\label{TAB1}
	\begin{tabular}{|l|c|c|c|c|}
		\hline
		& VBMax (eV) & CBMin (eV) & $m^*_h/m_e$ & $m^*_e/m_e$ \\
		\hline \hline
		AFM & -0.333 & 0.678 & 0.15 & 0.18 \\
		FM  & -0.339 & 0.555 & 0.02 & 0.11 \\
		\hline
	\end{tabular}
\end{table}

\begin{figure}[t!]
	\centering
	\includegraphics[clip=true, width=0.5\textwidth] {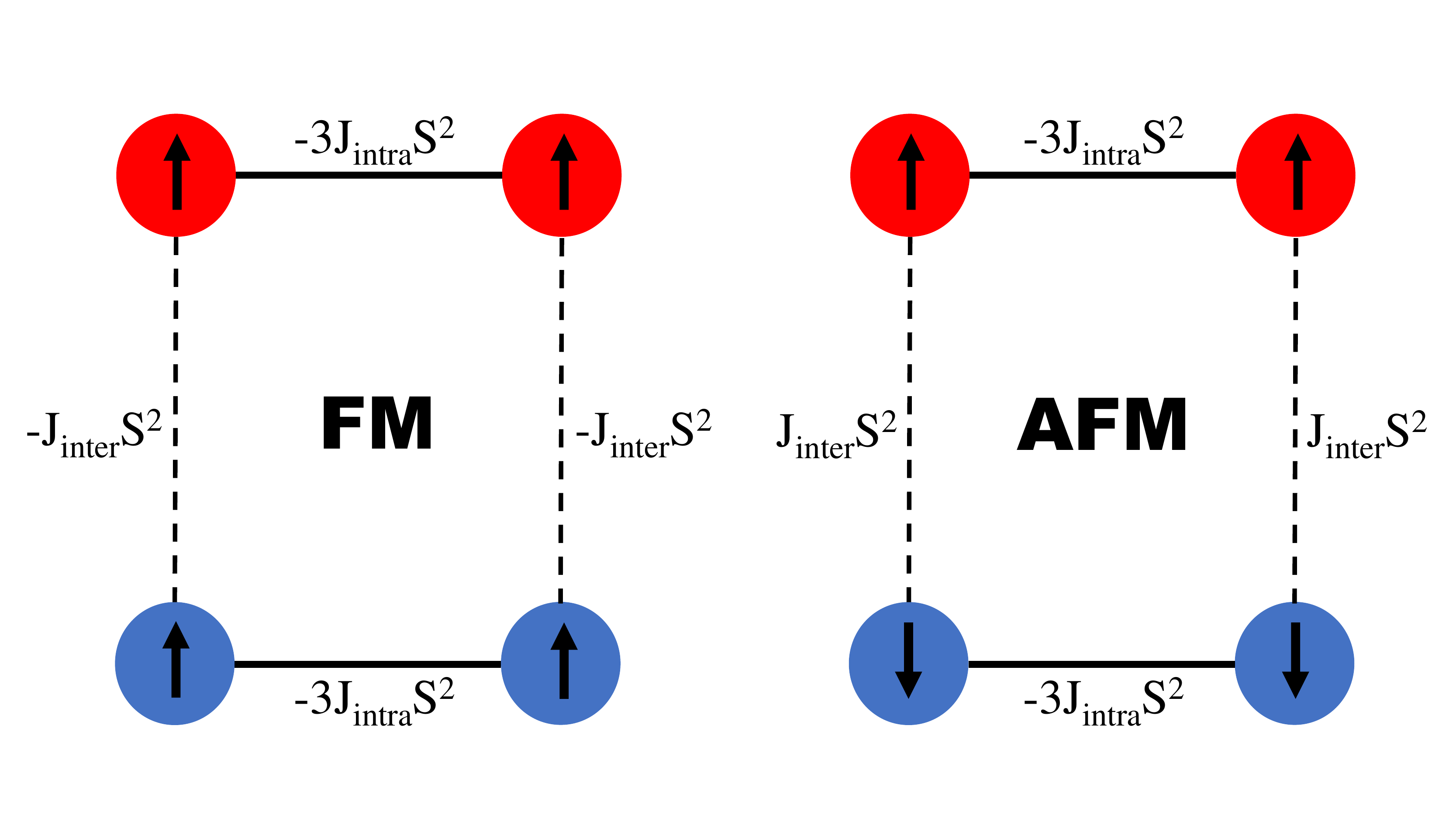}
	\caption{\label{FIG4} Schematic diagram describing inter- and intralayer exchange coupling in monoclinic bilayer CrI$_3$ for FM (left) and AFM (right) magnetic configurations. Top and bottom circles represent Cr atoms in different layers. The exchange energy ($E_{AFM} - E_{FM}$) per Cr atom is represented along the bonds connecting different magnetic atoms in the unit cell.}
\end{figure}

 The calculation of the interlayer exchange coupling $J_{inter}$ from first-principles calculations using a semiclassical approach\cite{LadoRossier2017} entails several difficulties due to the complexity of the Cr-Cr interlayer hopping at the microscopic level.\cite{SorianoRossier2019} Here, we adopt a simplified version by considering only a one-to-one Cr-Cr interlayer hopping (Fig.\ref{FIG4}). By doing so, we can write inter- and intralayer exchange couplings in terms of the total energies obtained using first-principles as:
\begin{eqnarray}
E_{FM}  & = & -6J_{intra}S^2 -2J_{inter}S^2 \nonumber \\
E_{AFM} & = & -6J_{intra}S^2 +2J_{inter}S^2 \nonumber \\
E_{AFM} - E_{FM} & = & 4J_{inter}S^2
\label{eqn1}
\end{eqnarray}
where $E_{AFM} - E_{FM}=\Delta E(n)$ is the interlayer exchange energy, which depends on the electrostatic doping $n$, $J_{intra}$ is the intralayer exchange coupling, and $S=3/2$ is the total spin per Cr atom. In the ground state ($n=0$), our first-principles calculations predict an interlayer AFM ground state with $\Delta E(0) = -75$ $\mu $eV/Cr ($J_{inter} = -8.1$ $\mu$eV/Cr).

\begin{figure}[t]
	\centering
	\includegraphics[clip=true, width=0.48\textwidth] {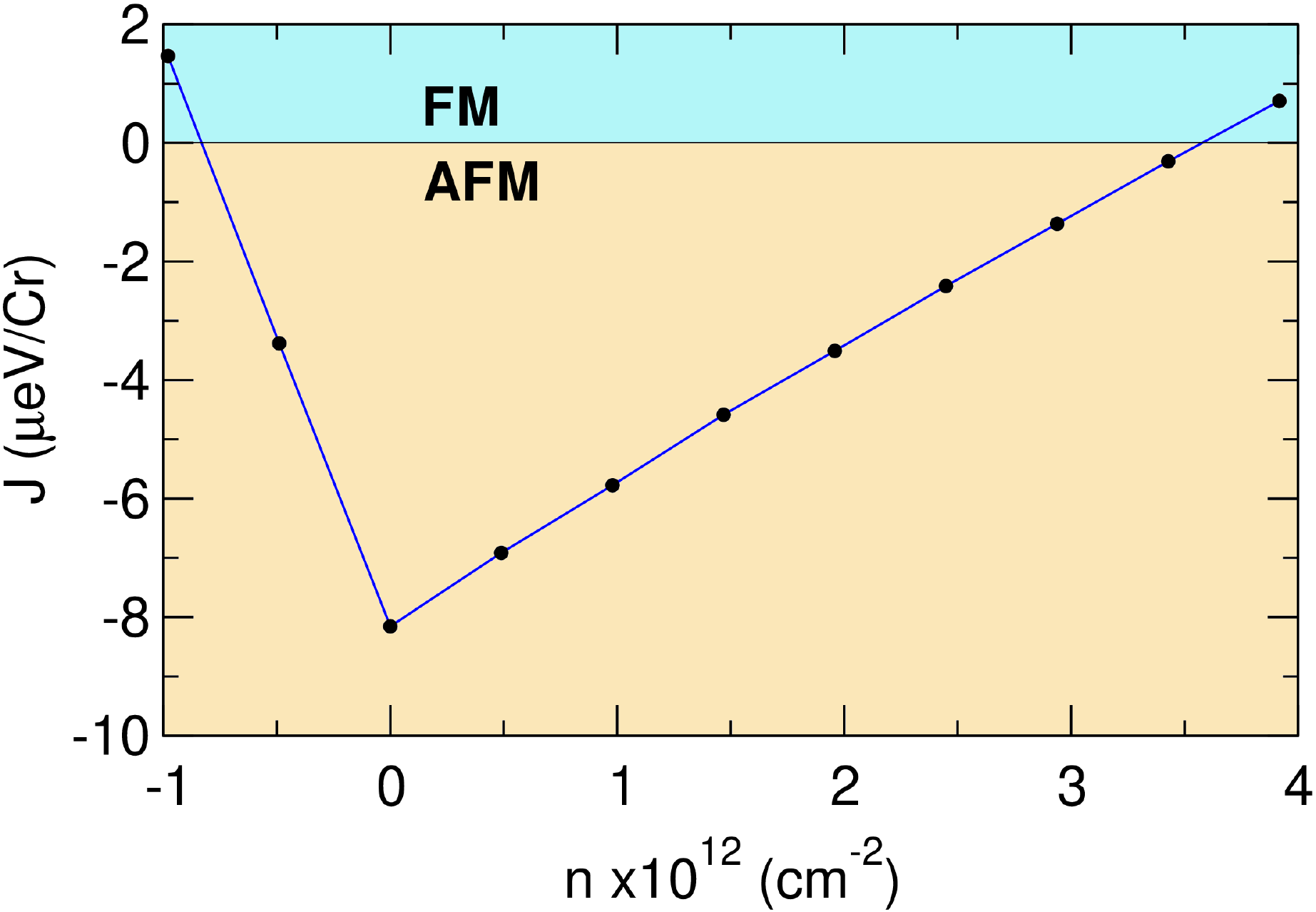}
	\caption{\label{FIG5} Interlayer exchange coupling dependece of electrostatic doping. Blue area belongs to the ferromagnetic transition.}
\end{figure}

Now, we study the effect of electrostatic doping on the interlayer exchange coupling in bilayer CrI$_3$ ($J_{inter}$). In Fig.\ref{FIG5}, we show the interlayer exchange coupling dependence on the electron and hole doping using first-principles calculations. The most interesting feature of this figure is the pronounce asymmetry around $n=0$. The blue and yellow regions correspond to FM and AFM ground states. The doping needed to switch the interlayer magnetism is around 4 times higher for holes than for electrons, namely $n_p/|n_e| \approx 4$ for the ferromagnetic transition. This has already been observed in a recent experiment by Huang {\it et al.}\cite{HuangXu2018}

\smallskip

{\textbf{\textit{Effective model.}}} Let us now discuss qualitatively why doping transforms antiferromagnetic ordering into ferromagnetic. In general, a weak doping {\it always} favors ferromagnetism. Indeed, in ferromagnetic environment electron propagates with a given direction of spin with maximally possible hopping whereas in antiferromagnetic or paramagnetic cases it has to arrange its spin to the direction of localized magnetic moment which leads to effective narrowing of the gap and increases its average band energy, the mechanism known as double exchange \cite{anderson,BR,AK6,nagaev1}. If the concentration of electrons is large enough, this ferromagnetic double exchange overcomes the initial antiferromagnetic exchange and the system turns into a ferromagnet, exactly as we see in our calculation. If the concentration of electrons is not high enough for this, the electron is supposed to form around itself a local ferromagnetic environment where it gets self-trapped, a phenomenon known as magnetic polaron, fluctuon, or ferron \cite{nagaev1,nagaev2}\cite{krivoglaz1,*krivoglaz2}\cite{AK1,*AK2}\cite{AK3,AK4,AK5}. When the concentration of electrons or holes increases, the system is phase separated into ferromagnetic droplets containing all charge carriers within an antiferromagnetic insulating matrix. This phase separation was found numerically within the narrow-band Hubbard model by Visscher \cite{visscher}; its formal theory was developed for the Hubbard and s-d exchange models at the Bethe lattice in Ref. \cite{AK6}.

Here, to have a physically transparent and quantitatively correct estimate, we will follow a simplified description of the magnetic polaron suggested by Mott \cite{mott}. We will assume that the electron (or hole) is self-locked in a radially symmetric ferromagnetic region considering a potential well with impenetrable walls. We start with the fact that, according to our calculations, ferromagnetic ordering between the layers, in comparison with the antiferromagnetic one, shifts the bottom of the conduction band down and the top of the valence band up, which is the driving force of the magnetic polaron formation and electron self-localization.

The total energy of the magnetic polaron in a layered ferromagnetic material can be written as
\begin{equation}
\mathcal{E}(R) = -\Delta + \frac{\hbar^2 z_0^2}{2m^*R^2} + J\frac{\pi R^2}{S_0},
\label{eqn2}
\end{equation}
where the first term $\Delta$ is the carrier energy difference between parallel and anti-parallel interlayer magnetization, namely $\Delta_{e(h)} = |\epsilon_{\rm CB(VB)}^{\rm AFM} - \epsilon_{\rm CB(VB)}^{\rm FM}|$, where VB and CB stands for the top and bottom of valence and conduction bands respectively. The second term is the energy of a particle confined in a disc of radius $R$ (Fig.\ref{FIG1}), with $z_0 = 2.40483$ being the first zero of the Bessel function $J_0(z)$. The third term is the exchange energy needed to switch the magnetic interaction between adjacent layers, where $J$ is the interlayer exchange coupling per magnetic atom, $S_0 = 3a^2\sqrt{3}/4$ is the area occupied by a single Cr atom in the unit cell, and $a$ is the Cr-Cr intralayer distance.

The minimum of the total energy in Eq.\ref{eqn2} with respect to the radius $R$ is
\begin{equation}
\mathcal{E}(R^*) = -\Delta + 2\sqrt{\frac{\hbar^2z_0^2}{2m^*}\frac{J\pi}{S_0}}
\label{eqn3}
\end{equation}
with optimal $R^* = \sqrt{(\hbar^2z_0^2/2m^*) (s_0/(J\pi))}$.

Then, the $R^*/a$ ratio is given by
\begin{equation}
\frac{R^*}{a} = \left(\frac{W}{J}\frac{z_0^2 3\sqrt{3}}{4\pi}\right)^{\frac{1}{4}}
\label{eqn4}
\end{equation}
where $W = \hbar^2/(2m^*a^2)$ is the effective carrier bandwidth. The model is applicable assuming that this ratio is much larger than one, which is necessary for the continuum-medium description adopted by us. As we will see further, this condition is satisfied in both our cases, for electrons and for holes.

Now, we rewrite Eq.\ref{eqn1} in terms of the carrier energy difference ($\Delta$), the interlayer exchange coupling ($J$) and the carrier bandwidth ($W$) as
\begin{equation}
\mathcal{E} = -\Delta + 2\sqrt{WJ\frac{4\pi z_0^2}{3\sqrt{3}}}.
\label{eqn5}
\end{equation}

By using Eqs.\ref{eqn4} and \ref{eqn5}, together with the  parameters obtained from first-principles calculations, we can study the feasibility of magnetic polaron formation in bilayer CrI$_3$. First, we need to calculate the effective carrier bandwidths
\begin{equation}
W = \frac{\hbar^2}{2m^*a^2} = {\rm Ry} \left( \frac{a_B}{a}\right)^2 \frac{m_e}{m^*}
\label{eqn6}
\end{equation}
where ${\rm Ry} = \hbar^2/(2m_ea_B^2) = 13.6$ eV, $a_B =  0.529$ \AA~ is the Bohr atomic radius and $m_e$ is the free-electron mass. Using Eq.\ref{eqn6}, we obtain $W_e = 2.2$ eV and $W_h = 12$ eV for electrons and holes, respectively. Now, by applying Eq.\ref{eqn4}, we obtain $(R^*_e/a) = 28.4$ and $(R^*_h/a) = 43.4$ which is more than enough to justify our simplified continuous-medium description.

Finally, we calculate the ground state energy of the magnetic polaron for both carriers using Eq.\ref{eqn5}. From the first two columns of Table \ref{TAB1}, we can calculate the energy difference for electron and hole carriers between the FM and AFM magnetic configurations,  $\Delta_e \approx 123$ meV and $\Delta_h \approx 6$ meV. The total energies are $\mathcal{E}_e = -91.3$ meV and $\mathcal{E}_h = 67.9$ meV, which means that, within our approximation, the formation of a magnetic polaron with interlayer FM ordering in bilayer CrI$_3$ is possible only for electrons. 

In summary, we have predicted the formation of a magnetic polaron in electron-doped bilayer CrI$_3$ using the parameters taken from first-principles calculations. These calculations also show that for electron doping the system transforms from the antiferromagnetic interlayer coupling to the ferromagnetic one. Our results give a possible explanation for the electrostatic interlayer magnetic switching experiments reported so far. It would be very interesting to study electron mobility for the case of very weakly doped system. For the case of electrons, it should be extremely small, due to electron self-trapping, and there should be a dramatic difference between electron- and hole-doped cases. The other prediction is that for the case of electron doping the atiferro-to-ferro transition should go via the intermediate two-phase region, whereas for the case of holes, due to the lack of magnetic polarons, there is no reason to expect this intermediate state

\smallskip

{\textbf{\textit{Acknowledgments.}}} We  acknowledge Efr\'en Navarro Moratalla and Malte R\"osner for fruitful discussions. D.S. thanks financial support from EU through the MSCA project Nr. 796795 SOT-2DvdW. M.I.K. acknowledges finacial support by JTC-FLAGERA Project GRANSPORT.

\bibliography{biblio}

\end{document}